\documentclass[11pt]{article}
\usepackage{fleqn,cospar,natbib}

\usepackage{url}

\usepackage{graphicx}

\newcommand{\apj}{{Astrophys. J.}}
\newcommand{\apjs}{{Astrophys. J. Supp.}}
\newcommand{\apjl}{{Astrophys. J. Lett.}}
\newcommand{\aj}{{Astron. J.}}
\newcommand{\mnras}{{Mon. Not. R. Astron. Soc.}}
\newcommand{\aap}{{Astron. \& Astrophys.}}

\newcommand{\nat}{{Nature}}

\title{X-ray Spectroscopy and Variability of AGN
Detected in the 2~Ms Chandra Deep Field-North Survey}

\author{F.E.~Bauer,\address{Department of Astronomy \& Astrophysics, 525 Davey Lab, 
The Pennsylvania State University, University Park, PA 16802}
C.~Vignali,$^{1}$ D.M.~Alexander,$^{1}$ W.N.~Brandt,$^{1}$
G.P.~Garmire,$^{1}$ A.E.~Hornschemeier,$^{1}$ P.S.~Broos,$^{1}$
L.K.~Townsley,$^{1}$ and D.P.~Schneider$^{1}$}

\begin{document}

\maketitle

\begin{abstract}
We investigate the nature of the faint X-ray source population through
X-ray spectroscopy and variability analyses of 136 AGN detected in the
2 Ms Chandra Deep Field-North survey with $> 200$
background-subtracted 0.5--8.0~keV counts [$F_{\rm
0.5-8.0~keV}=(1.4$--$200)\times10^{-15}$~erg~cm$^{-2}$~s$^{-1}$]. Our
preliminary spectral analyses yield median spectral parameters of
$\Gamma=1.61$ and intrinsic $N_{\rm H}=6.2\times10^{21}$~cm$^{-2}$
($z=1$ assumed when no redshift available) when the AGN spectra are
fitted with a simple absorbed power-law model. However, considerable
spectral complexity is apparent (e.g., reflection, partial covering)
and must be taken into account to model the data accurately. Moreover,
the choice of spectral model (i.e., free vs. fixed photon index) has a
pronounced effect on the derived $N_{\rm H}$ distribution and, to a
lesser extent, the X-ray luminosity distribution. Ten of the 136 AGN
($\approx7$\%) show significant Fe K$\alpha$ emission-line features
with equivalent widths in the range 0.1--1.3~keV. Two of these
emission-line AGN could potentially be Compton thick (i.e., $\Gamma <
1.0$ and large Fe K$\alpha$ equivalent width). Finally, we find that
81 ($\approx60$\%) of the 136 AGN show signs of variability, and that
this fraction increases significantly ($\approx80$--90\%) when better
photon statistics are available.
\end{abstract}

\section*{INTRODUCTION}

With the bulk of the $\approx$~0.5--10.0 keV background now resolved
into discrete point sources \citep[e.g.,][]{Mushotzky2000,
Brandt2001b, Rosati2002}, emphasis has shifted toward determining the
nature of the faint X-ray population --- a clear necessity if we wish
to understand the formation and evolution of active galactic nuclei
(AGN). X-ray spectroscopy and variability analyses can play a crucial
role in constraining the physical processes which occur in these AGN. 
For instance, the numerous X-ray spectroscopic and variability studies
made prior to the launches of {\it Chandra} and {\it XMM-Newton} have
found that
(1) AGN X-ray spectra often show signs of complex absorption,
reflection, and emission lines in addition to their power-law
continuum radiation \citep[e.g.,][]{
Nandra1994,
Smith1996, 
Lawson1997,
Nandra1997c, 
Turner1997,
George2000,
Reeves2000},
(2) the strengths of these spectral features (especially the
reflection and emission components) depend on luminosity
\citep[e.g.,][]{
Iwasawa1993,
Nandra1997a},
and (3) a large fraction of AGN demonstrate variability of either
continuum emission or absorption
\citep[e.g.,][]{
Nandra1997b,
Turner1999,
Almaini2000,
Grupe2001,
Manners2002,
Risaliti2002a}. However, these studies have generally focused on small
samples of bright objects ($>1\times10^{-13}$~erg~cm$^{-2}$~s$^{-1}$),
which are comprised of either nearby, moderate-luminosity AGN or rare,
high-luminosity AGN at large look-back times. The deepest observations
performed by {\it Chandra} and {\it XMM-Newton} now present the
opportunity to study AGN $\sim$~100 times fainter in similar detail
and thus place constraints on moderate-luminosity AGN at earlier
epochs. Here we report on the spectral and temporal properties of the
AGN in the 2~Ms exposure of the {\it Chandra} Deep Field-North
(CDF-N), which probes $\approx2$--10 times fainter in the 0.5--8.0~keV
band than any other {\it Chandra} or {\it XMM-Newton} survey, and at
least 50 and 250 times fainter than any past X-ray instrument in the
0.5--2.0~keV and 2.0--8.0~keV bands, respectively.

\section*{DATA}

\begin{figure}
\vspace{-1.5in}
\parbox{8.0cm}{
\small\baselineskip 9pt
\centerline{
\includegraphics[width=8.0cm,angle=0]{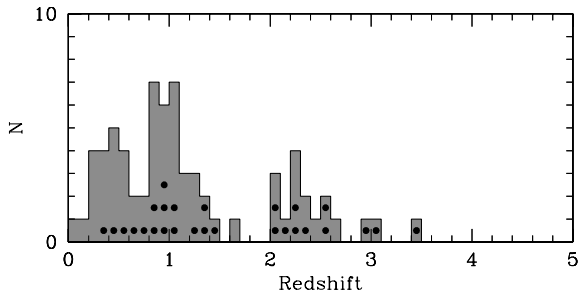}
}
\vspace{0.0cm} 
}
\hfill
\parbox{8.0cm}{
\vspace{2.0cm} 

\caption{Redshift distribution for the
72 sources in the sample with spectroscopic redshifts. Black dots
indicate the sources which are confirmed broad-line AGN
\citep{Cohen2000,Barger2002}.
\label{fig:zdist}}
}
\vspace{-0.5cm} 
\end{figure}

Our sample was selected to ensure that we have adequate photon
statistics for spectral and temporal analyses and is comprised of the
136 extragalactic X-ray sources with more than 200
background-subtracted, 0.5--8.0~keV counts in the 2~Ms CDF-N catalog
\citep[out of 503 total;][]{Alexander2003b}; X-ray spectral analyses of 
a few fainter X-ray sources are presented in \citet{Vignali2002} and
\citet{Alexander2003a}. With 0.5--8.0~keV fluxes ranging
from $1.4\times10^{-15}$ to $2\times10^{-13}$~erg~cm$^{-2}$~s$^{-1}$,
these sources are characteristic of the population which comprises the
bulk of the X-ray background below 10~keV. Seventy-two
($\approx$~53\%) of the sources in the sample have spectroscopic
redshifts (see Figure~\ref{fig:zdist}); among these, nearly all
have X-ray luminosities consistent with AGN. Based on their spectral
slopes and X-ray-to-optical flux ratios (i.e., most have $I>23$), the
remaining unidentified sources are likely to be optically faint,
obscured AGN at $z \sim 1$--3 \citep[e.g.,][]{Alexander2001,
Barger2001a}. Thus almost all of the 136 sources appear to be AGN. 
Since the unidentified sources comprise almost half of the sample and
may play an important role in the overall intrinsic column density and
luminosity distributions, we have placed the unidentified sources at
$z=1$ (i.e., the likely lower redshift bound for these sources).

\section*{SPECTRAL ANALYSIS}

\begin{figure}[bp]
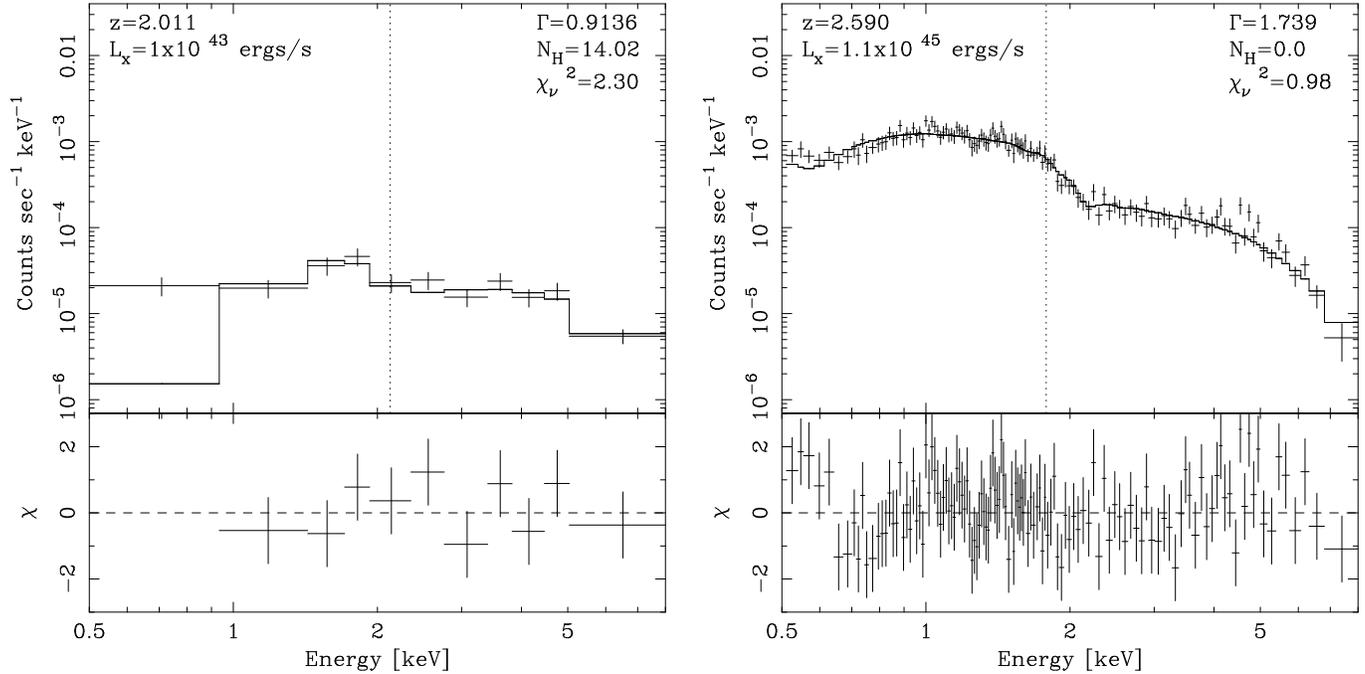

\vspace{0.0in}
\parbox{7.9cm}{
\small\baselineskip 9pt
\centerline{
\hglue0.3in{\includegraphics[width=8.9cm,angle=-90]{123635.6+621424.spectra.ps}}
}
}
\hfill
\parbox{7.9cm}{ 
\small\baselineskip 9pt
\centerline{
\hglue-0.3in{\includegraphics[width=8.9cm,angle=-90]{123622.9+621527.spectra.ps}}
}
}
\vspace{-0.7cm} 
\caption{Two example X-ray spectra fit with MODEL~1. The top panel shows 
a model fit to the data plotted in the observed-frame energy, while
the bottom panel shows the residuals plotted in terms of $\chi$. The
relevant spectral fit parameters are shown inset ($N_{\rm H}$
is given in units of $10^{20}$~cm$^{-2}$). The vertical dotted line
indicates the energy of the 6.4~keV Fe~K$\alpha$ line at the source
redshift. {\bf Left:} CXOHDFN~123635.6$+$621424, an X-ray obscured
Seyfert~2 galaxy
\citep[see also][]{Dawson2001}. {\bf Right:}
CXOHDFN~123622.9$+$621527, an unobscured quasar.
\label{fig:example_spectra}}
\vspace{-0.5cm} 
\end{figure}

\begin{figure}
\parbox{8.5cm}{
\small\baselineskip 9pt
\vspace{-1.0cm} 
\centerline{
\includegraphics[width=8.4cm]{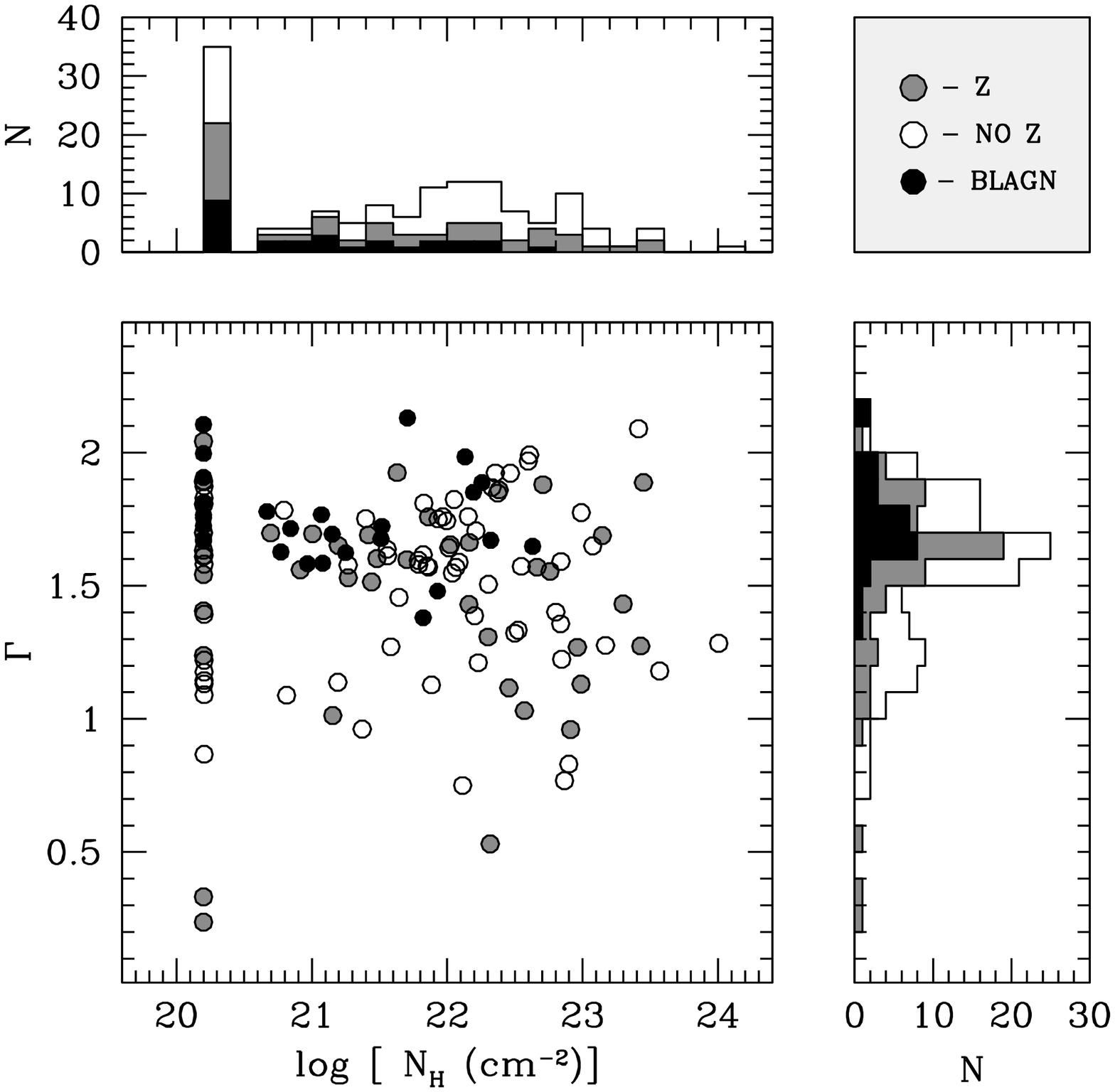}
}
\vspace{-0.7cm} 
\caption{Intrinsic $N_{\rm H}$ versus $\Gamma$ (and their
distributions) for the 136 X-ray sources in our 
sample fit with MODEL~1. Sources without redshifts are taken to have $z=1$.
\label{fig:nh_vs_gamma}}
}
\hfill
\parbox{8.5cm}{
\small\baselineskip 9pt
\vspace{-0.3cm} 
\centerline{
\includegraphics[width=8.65cm]{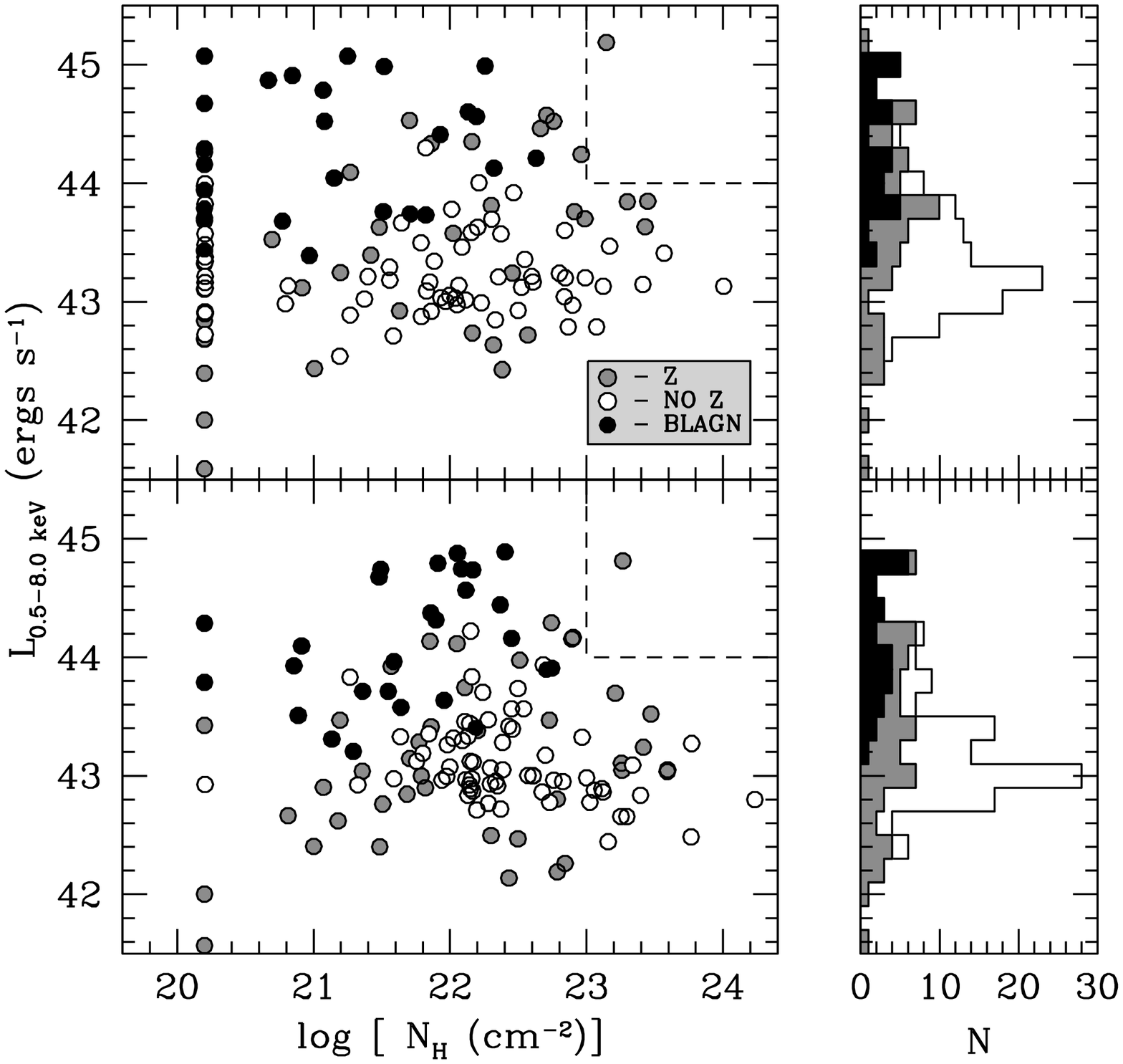}
}
\vspace{-0.8cm} 
\caption{Intrinsic $N_{\rm H}$ versus $L_{\rm X}$ for the 136 
X-ray sources in our sample fit with MODEL~1 ({\bf top})
and MODEL~2 ({\bf bottom}). Sources without redshifts are taken to have
$z=1$. The dashed lines indicate the approximate region where obscured
quasars should lie.
\label{fig:nh_vs_Lx}}
}
\end{figure}

Details of our spectral analysis procedure are given in
\citet{Bauer2003}. X-ray spectral fitting was performed using XSPEC
\citep[v11.2;][]{Arnaud1996}. We initially fitted each source with a
simple model (hereafter MODEL~1) consisting of fixed Galactic
photoelectric absorption ($N_{\rm H_{GAL}}=1.6\times10^{20}$~cm$^{-2}$
toward the CDF-N), a varying intrinsic $N_{\rm H}$ (using the redshift
if known), and a power law with varying photon index
($\Gamma$). Approximately 75\% of the sources are acceptably fit by
MODEL~1 [$P(\chi^{2}|\nu) < 0.05$]. Example spectra of two typical
\hbox{CDF-N} AGN are shown in Figure~\ref{fig:example_spectra}. The
MODEL~1 fits yield median spectral parameters of $\Gamma=1.61$ and
$N_{\rm H}=6.2\times10^{21}$~cm$^{-2}$ for our 136 sources, with large
dispersions for both parameters; the overall distributions are shown
in Figure~\ref{fig:nh_vs_gamma}. Note that the changes in these
parameters are much less than their dispersions if we limit ourselves
to the sources with redshifts (i.e., $\Gamma=1.67$ and $N_{\rm
H}=1.8\times10^{21}$~cm$^{-2}$) or with $>$ 500 counts (i.e.,
$\Gamma=1.69$ and $N_{\rm H}=3.2\times10^{21}$~cm$^{-2}$). The $N_{\rm
H}$ distribution has a peak at $N_{\rm H}\approx$~Galactic, and very
few objects ($<9$\%) have $N_{\rm H} > 10^{23}$~cm$^{-2}$. The 27
optically identified broad-line AGN (BLAGN) appear to trace the same
$N_{\rm H}$ distribution as the overall sample. The typical photon
indices found for the CDF-N AGN are $\Delta \Gamma \sim 0.2$--0.3
lower than the canonical intrinsic X-ray spectral slope value of AGN
\citep[e.g., $\Gamma \sim 1.9$--2.0;][]{Nandra1994, Brandt1997,
Reeves2000}, suggesting the presence of additional spectral complexity
(e.g., reflection, complex intrinsic absorption, partial covering). 
This fact is strengthened somewhat by the systematic residuals (e.g.,
soft excesses, spectral curvature) seen in several objects. 

Detailed modeling of this complexity is underway but is beyond the
scope of this paper. As a simple test of how the $N_{\rm H}$ and
$L_{\rm X}$ distributions could be affected by spectral complexity, we
fitted the data with a model identical to MODEL~1, but with the photon
index fixed to $\Gamma=2$ (i.e., close to the canonical intrinsic
slope found locally; hereafter MODEL~2). We find that fewer sources
are acceptably fit [$P(\chi^{2}|\nu) < 0.05$] with MODEL~2
($\approx$~60\%) and many sources have strong systematic residuals;
this again suggests that there is spectral complexity. The median
intrinsic absorption is $N_{\rm H}=1.5\times10^{22}$~cm$^{-2}$ and the
median differences between intrinsic $N_{\rm H}$ and $L_{\rm X}$ for
MODEL~1 and MODEL~2 are factors of 2.5 and 0.5, respectively (see
Figure~\ref{fig:nh_vs_Lx}). The decrease in X-ray luminosity found for
MODEL~2 is due in part to $K$-correction differences from the free
versus fixed photon indices; note that the intrinsic X-ray luminosity
could increase substantially if reflection and scattering were
actually taken into account.

\section*{EMISSION LINES}

\begin{figure}
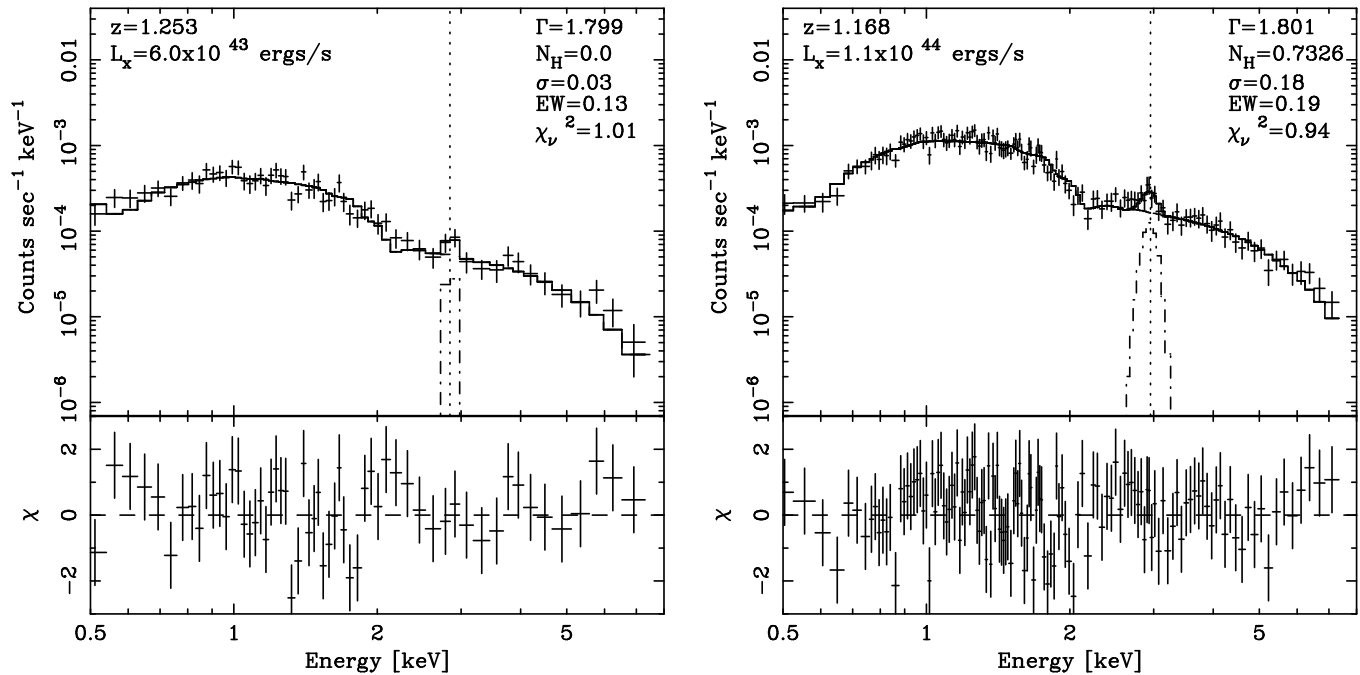

\vspace{0.0in}
\parbox{7.9cm}{ 
\small\baselineskip 9pt
\centerline{
\hglue0.3in{\includegraphics[width=8.9cm,angle=-90]{123704.1+620756.spectra.ps}}
}
}
\hfill
\parbox{7.9cm}{ 
\small\baselineskip 9pt
\centerline{
\hglue-0.3in{\includegraphics[width=8.9cm,angle=-90]{123740.9+621201.spectra.ps}}
}
}
\vspace{-0.7cm} 
\caption{Two example emission-line X-ray spectra fit with MODEL~1 plus 
a Gaussian emission line. See the Figure~\ref{fig:example_spectra} caption
for explanation (note $N_{\rm H}$, $\sigma$, and EW are given in
units of $10^{20}$~cm$^{-2}$, keV, and keV, respectively). 
{\bf Left:} CXOHDFN~123704.1$+$620756, a BLAGN with a narrow emission
line. {\bf Right:} CXOHDFN~123740.9$+$621201, an obscured AGN with a
mildly broadened or complex emission line.
\label{fig:line_example_spectra}}
\vspace{-0.5cm} 
\end{figure}

X-ray background synthesis models and X-ray spectral analyses of
nearby sources both suggest that a large fraction of AGN may be
Compton thick \citep[e.g.,][]{Comastri1995, Risaliti1999}, with direct
emission observable only above rest-frame energies of $\sim$~10~keV
(if at all). Below this energy, these AGN can only be detected via
their faint reflected or scattered emission, typically in the form of
extremely flat spectral slopes and large equivalent-width (EW)
emission lines \citep[e.g.,][]{Maiolino1998, Matt2000}. We can place
constraints on the existence of such objects at faint X-ray fluxes by
searching for these features in our sample. For instance, 10 of the
136 AGN ($\approx$7\%) exhibit obvious Fe~K$\alpha$ emission-line
features with EW~$=$~0.1--1.3~keV. The emission lines appear to be
typically a combination of narrow 6.4, 6.7, and 6.96~keV emission
lines, although a few objects exhibit a possible broad wing redward of
the 6.4~keV line. Example spectra of two typical CDF-N emission-line
AGN are shown in Figure~\ref{fig:line_example_spectra}. Two of the
emission-line sources are potentially Compton-thick AGN with $\Gamma <
1.0$ and EW~$=$~0.6--0.7~keV. We can place further constraints on the
number of reflection-dominated Compton-thick AGN in two ways. First,
only 11 of the 136 sources have measured photon indices $\Gamma <
1.0$, suggesting that few objects in the sample show the signature of
pure reflection that is characteristic of reflection-dominated
Compton-thick AGN. Second, by modeling a Gaussian component to the
source spectrum ($E=6.4$~keV, $\sigma=0$~keV), we can place upper
limits on the Fe K$\alpha$ EWs for the 72 objects with redshifts. The
range of Fe K$\alpha$ EW upper limits is 0.1--2.9~keV \citep[i.e.,
consistent with local Seyferts; e.g.,][]{Nandra1994}, indicating that
many sources may have emission lines which are still individually
undetectable with the current data. However, only 8 of the 72 sources
($\approx$11\%) have 90\% confidence EW upper limits above 1~keV, and,
of these, none has a measured photon index $\Gamma < 1.0$. Given that
$\sim$50\% of all known Compton-thick sources have $\Gamma < 1.0$ and
EW~$>$~1~keV
\citep[e.g.,][]{Maiolino1998, Bassani1999}, and that large EWs can
also arise from anisotropic ionizing radiation or lags between
continuum and line variability, the true number of Compton-thick
AGN among our sample is likely to be small.

\section*{VARIABILITY}

The 20 individual CDF-N observations span $\approx27$ months and offer
an unprecedented probe of X-ray variability in distant AGN. Using
Kolmogorov-Smirnov (single observations, second-to-day timescales) and
$\chi^{2}$ (multiple observations, day-to-year timescales) tests, we
find that $\sim$~55\% and $\sim$~60\% of the 136 sources are variable
at greater than 99\% confidence, respectively. These fractions
increase significantly when we restrict ourselves to the 61 objects
with $> 500$ counts (i.e., $\sim$~80\% and $\sim$~90\%, respectively)
or the 27 objects with broad lines (i.e., $\sim$~66\% and $\sim$~92\%,
respectively). The latter results are most likely a result of better
photon statistics alone, suggesting that the vast majority of the
CDF-N AGN are variable. We detect maximum and median variability
amplitudes in the vignetting-corrected count rate of $\approx$~7.8 and
$\approx$~1.9, respectively, with 3 (15) sources having maximum
variability amplitudes larger than 5 (3). The variability properties
of these moderate-luminosity AGN are consistent with local studies of
Seyfert galaxies
\citep[e.g.,][]{Nandra1997b}.

\begin{figure}
\vspace{-0.1in}
\parbox{5.5cm}{ 
\small\baselineskip 9pt
\centerline{
\hglue0.15in{\includegraphics[width=5.8cm,angle=0]{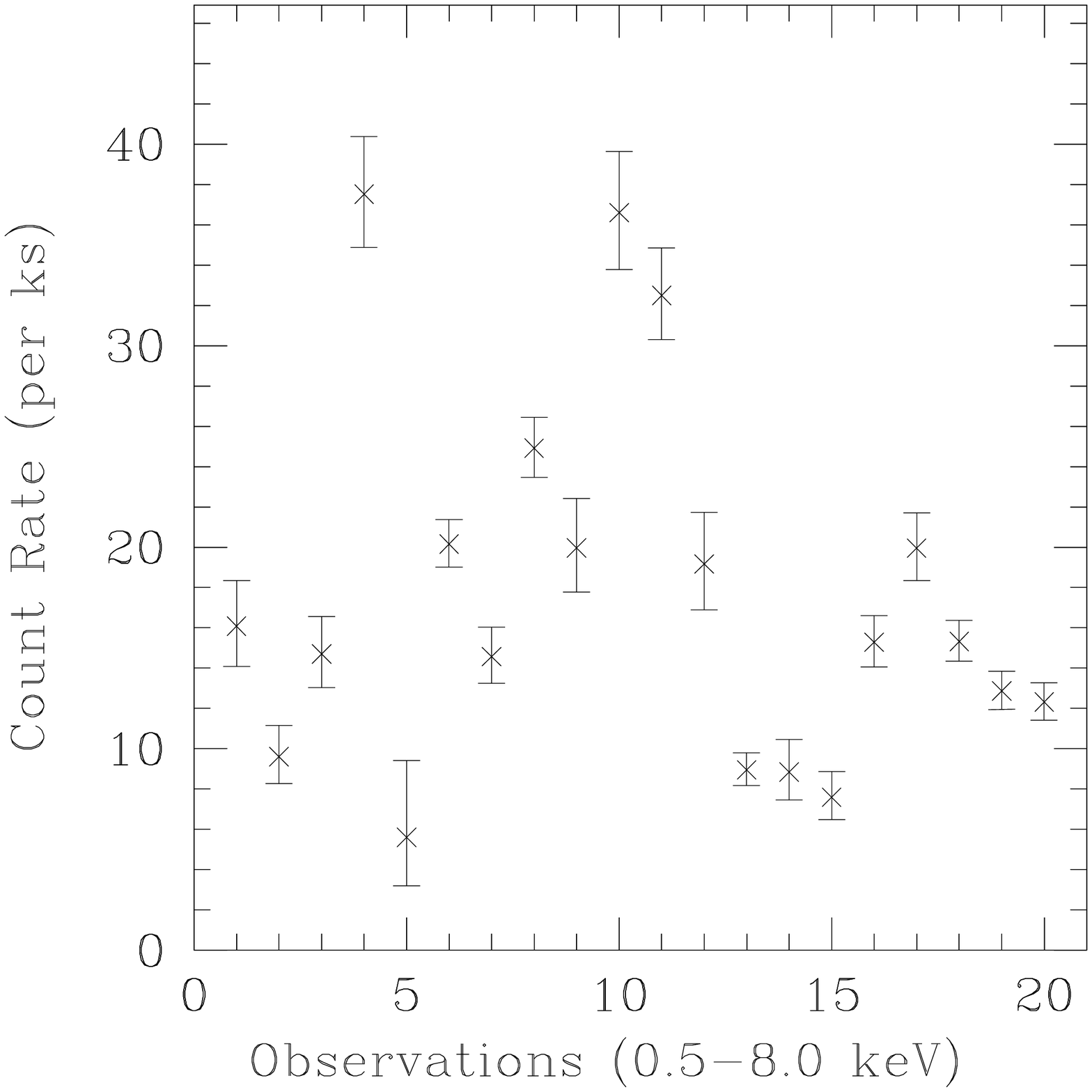}}
}
}
\hfill
\parbox{5.5cm}{ 
\small\baselineskip 9pt
\centerline{
\hglue0.05in{\includegraphics[width=5.8cm,angle=0]{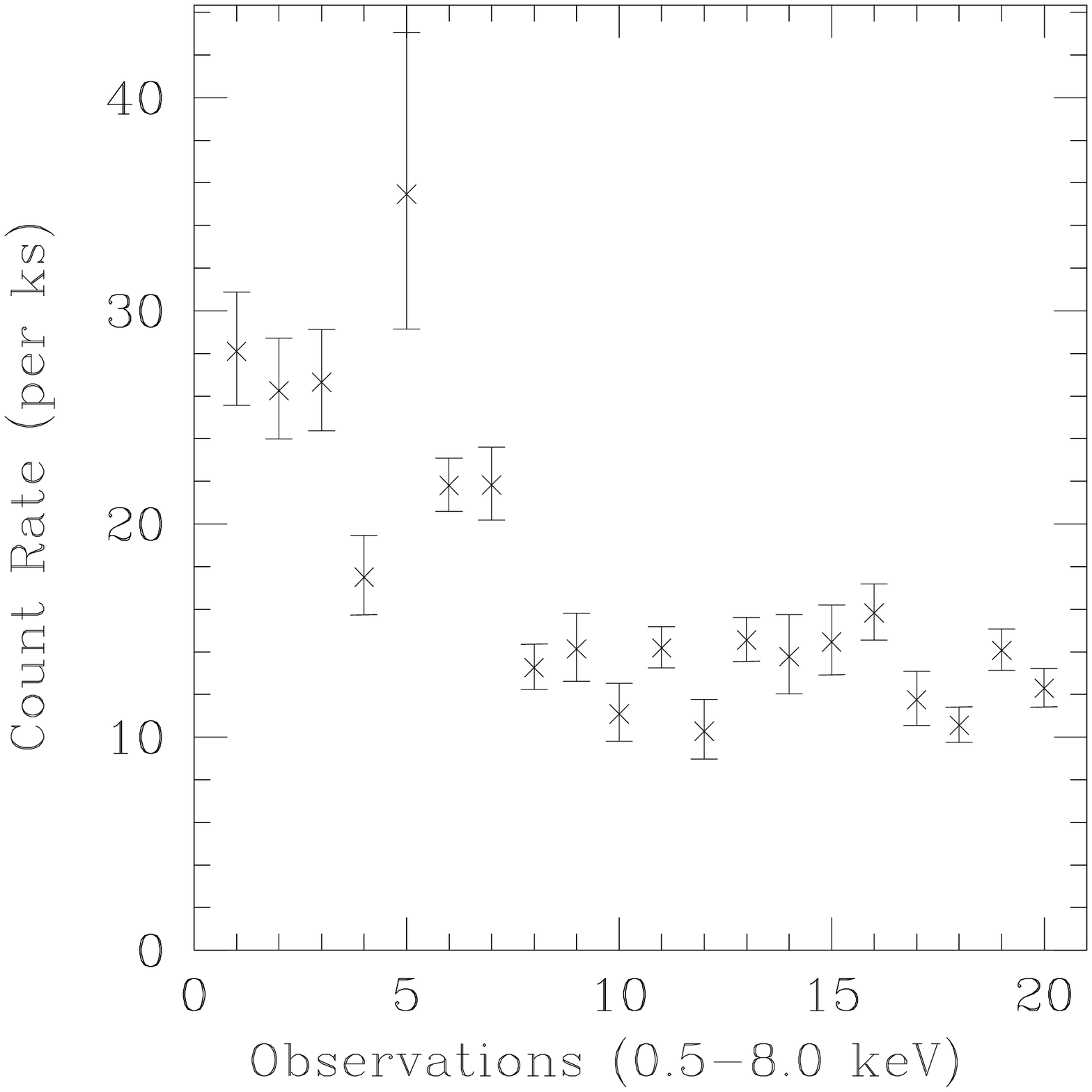}}
}
}
\hfill
\parbox{5.5cm}{ 
\small\baselineskip 9pt
\centerline{
\hglue0.0in{\includegraphics[width=5.8cm,angle=0]{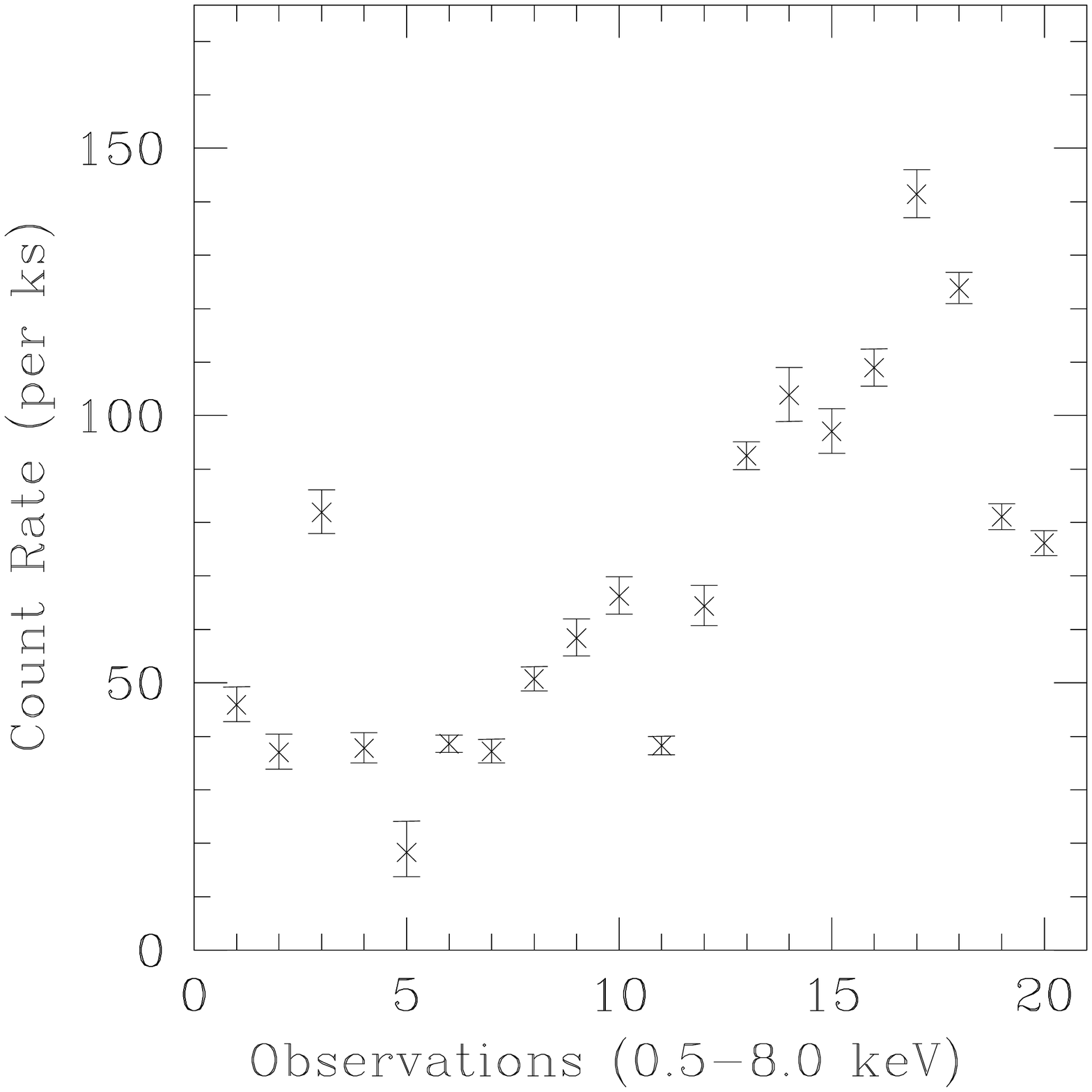}}
}
}
\vspace{-0.5cm} 
\caption{Count rate (per ks) versus observation number 
for three example sources. Note that the count rates have been
corrected for vignetting. {\bf Left:} CXOHDFN~123612.0$+$621139, a
Seyfert~1.9 galaxy with $z=0.275$, $L_{\rm X}=4\times10^{42}$ erg~s$^{-1}$.
CXOHDFN~123618.0$+$621635, a narrow emission-line ({\sc [Oiii]} only) galaxy with $z=0.679$, 
$L_{\rm X}=4\times10^{43}$ erg~s$^{-1}$.
{\bf Right:}
CXOHDFN~123752.7$+$621628, a BLAGN with $z=0.307$, $L_{\rm X}=2\times10^{43}$ erg~s$^{-1}$.
\label{fig:var_fig}}
\end{figure}

\section*{ACKNOWLEDGEMENTS}

We gratefully acknowledge the financial support of NSF CAREER award
AST-9983783 (FEB, CV, DMA, WNB), NASA grant NAS 8-38252 (GPG, PI), 
NASA GSRP grant NGT5-50247 (AEH), and NSF grant AST-9900703 (DPS).
This work would not have been possible without the support of the
entire {\it Chandra} and ACIS teams.


\end{document}